\documentclass[11pt,aps,showpacs,floatfix,onecolumn,tightenlines]{revtex4}
\usepackage{epsfig}
\usepackage{psfig}
\usepackage{graphicx}
\usepackage{graphics}
\usepackage{xspace}
\usepackage{amssymb}
\usepackage{amsmath}
\usepackage{latexsym}
\usepackage{natbib}
\usepackage{mathrsfs}
\newcommand{\inieq}{\begin{eqnarray}}            
\newcommand{\fineq}{\end{eqnarray}}            
\newcommand{\ove}{\overline}                    
\newcommand{\diff}{{\rm\,d}}                    

\def\p{\mbox{\boldmath $p$}}

\def\q{\mbox{\boldmath $q$}}

\def\k{\mbox{\boldmath $k$}}

\begin{document}
\title{Neutral-current neutrino-nucleus quasielastic scattering} 

\author{Andrea Meucci} 
\author{Carlotta Giusti}
\author{Franco Davide Pacati }
\affiliation{Dipartimento di Fisica Nucleare e Teorica, 
Universit\`{a} di Pavia and \\
Istituto Nazionale di Fisica Nucleare, 
Sezione di Pavia, I-27100 Pavia, Italy}


\begin{abstract}
A relativistic distorted-wave impulse-approximation model to quasielastic
neutral-current neutrino-nucleus scattering is developed, using the 
relativistic mean
field theory for the bound state and taking into account the final state
interaction in the relativistic scattering state. 
Results for the neutrino (antineutrino) reaction on $^{12}$C target nucleus
are presented in an energy range between 500 and 1000 MeV. The sensitivity to
the strange quark content of the axial-vector form factor is discussed. 
\end{abstract}
\pacs{25.30.Pt  13.15.+g  24.10.Jv}


\maketitle

\section{Introduction}

The neutral-current scattering of neutrinos and antineutrinos on nucleons and
nuclei has gained in recent years a wide interest in order to determine the
structure of hadronic weak neutral current. In a measurement of the proton spin
structure function, the European Muon Collaboration \cite{emc} found out a 
disagreement with
the Ellis-Jaffe sum rule \cite{ellis}, when only up and down quark and 
antiquark contribution to the proton spin was taken into account, thus indicating 
that other flavors might contribute to the spin structure of the proton. A
measurement of neutrino (antineutrino)-proton elastic scattering at Brookhaven
National Laboratory (BNL) \cite{bnl} supported this result 
by reporting a non-zero value for the strange axial-vector form factor of the 
nucleon. However, the BNL data were critically reanalyzed in Ref. \cite{gar},
where also strange vector form factors were taken into account. It turned out that the BNL 
data cannot provide us
decisive conclusions about the strange form factors \cite{gar}.
The BooNE experiment \cite{mini} at Fermi National Laboratory (FermiLab) 
aims,
through the FermiLab Intense Neutrino Scattering Scintillator Experiment 
(FINeSSE) \cite{fin}, 
at performing a detailed investigation of the strangeness contribution to the 
proton spin via neutral current elastic scattering. 

In order to reach a definite conclusion, nuclear structure effects have to be 
clearly understood. Quasielastic
electron scattering calculations \cite{book}, which were able to successfully 
describe a
wide number of experimental data, can provide us a useful tool to study
neutrino-induced processes. However, a careful analysis of neutrino-nucleus
reactions introduces additional complications, as the final neutrino cannot
in practice be measured and a final hadronic probe has to be detected: 
the corresponding cross
sections are therefore semi-inclusive in the hadronic sector and inclusive in 
the leptonic one. 

General review papers about neutrino-nucleus interactions can be found in Refs.
\cite{Walecka,peccei,alb,kolbe03}.
The relativistic Fermi gas (RFG) model was applied to study neutral-current
neutrino-induced knockout from nuclei in Ref. \cite{hor}, where binding energies
corrections and strange-quark axial-vector and vector effects were investigated.
Detailed analyses of nuclear structure effects on the determination of
strangeness contribution were performed in Refs. \cite{barbaro,alberico}. In
addition to the RFG model a {\lq\lq hybrid\rq\rq} model in which bound nucleons are
described by harmonic oscillator wave-functions was proposed in Ref.
\cite{barbaro}, while a relativistic shell model was used in Ref.
\cite{alberico}. The effects of final state interactions (FSI) were also studied
in Ref. \cite{gar93} in the framework of random phase approximation (RPA) and a 
continuum RPA 
model was developed in Ref. \cite{ryc}, where nucleosynthesis processes were
also discussed. 
The FSI effects are generally large on the cross section, but they are
reduced when studying the ratio of proton to neutron cross sections.
This was discussed in Refs. \cite{gar92,hor,alberico}. 
The effects of two-body relativistic meson exchange
currents in neutrino-nucleus scattering at low and intermediate energies were 
evaluated in Refs. \cite{umino,umino1}. The sensitivity of the neutrino-nucleus 
cross sections to the strange-quark contribution was also examined in a 
relativistic plane wave impulse approximation (RPWIA) in 
Ref. \cite{vdv}, where a model-independent
approach based on eight structure functions was developed. 

In this paper we present a relativistic distorted-wave impulse-approximation
(RDWIA) calculation of neutral-current $\nu$- and $\bar\nu$-nucleus reactions 
in the quasi\-elastic region, where a 
neutrino interacts with only one nucleon in the nucleus and the others 
nucleons behave as spectators. 
The RDWIA treatment is the same as in Refs.~\cite{meucci1,meucci2,meucci3},
where it was applied to electromagnetic knockout reactions and in  
Ref. \cite{cc}, where a relativistic Green's function approach was developed 
dealing with the quasielastic charged-current $\nu$-nucleus scattering.
The relativistic 
bound state wave-functions are solutions of a Dirac equation 
containing scalar and vector potentials obtained in the framework of the 
relativistic mean field theory \cite{adfx,lala}.
The FSI are taken into account through a relativistic
optical model potential \cite{chc,clark}.
The effective Pauli reduction has been adopted in this case and the
resulting Schr\"odinger-like equations are solved for each partial wave.

The formalism is outlined in Sec. \ref{sec.for}. Results are presented and
discussed in Sec. \ref{results}. Some conclusions are drawn in 
Sec. \ref{conc}.

\section{The formalism of the semi-inclusive quasielastic scattering}
\label{sec.for}

The neutral-current $\nu$($\bar\nu$)-nucleus cross section for the
semi-inclusive process can be 
written as the contraction between the lepton tensor and the hadron 
tensor, i.e.,
\begin{eqnarray}
\diff \sigma = \frac {G^2 } {2} \ 2\pi \
 L^{\mu\nu}\ W_{\mu\nu}\ \frac {\diff^3k} {(2\pi)^3} \ 
 \frac {\diff^3p_{\mathrm N}} {(2\pi)^3} \ ,
\label{eq.cs1}
\end{eqnarray}
where $G \simeq 1.16639 \times 10^{-11}$ MeV$^{-2}$ is the Fermi constant, and 
$k^\mu_i = (\varepsilon_i,\k_i)$, $k^\mu = (\varepsilon,\k)$ are 
the four-momentum of the incident and final neutrino or antineutrino,
respectively, and $\p_{\mathrm N}$ is the momentum of the emitted nucleon.

The lepton tensor has a similar structure as in electromagnetic knockout or
in charged-current $\nu$-nucleus scattering and can be
written as in Ref. \cite{book,cc,vdv}. After projecting 
into the initial and the final neutrino (antineutrino)  
states, it is decomposed into a 
symmetrical and an antisymmetrical component, i.e.,
\begin{eqnarray}
L^{\mu\nu} = \frac {2} {\varepsilon_i \varepsilon} 
\left[ l_S^{\mu\nu} \mp l_A^{\mu\nu} \right],
\label{eq.lt2}
\end{eqnarray}
with
\begin{eqnarray}
l_S^{\mu\nu} &=& k_i^\mu \ k^\nu + k_i^\nu \ k^\mu - g^{\mu\nu} \ k_i \cdot k
\nonumber \\
l_A^{\mu\nu} &=& i \ \epsilon ^{\mu\nu\alpha\beta} k_{i\alpha} k_\beta ,
\label{eq.lt3}
\end{eqnarray}
where $\epsilon ^{\mu\nu\alpha\beta}$ is the antisymmetric tensor with 
$\epsilon_{0123} = - \epsilon^{0123} = 1$.
The upper (lower) sign in Eq. \ref{eq.lt2} refers to $\nu$($\bar\nu$) 
scattering.

Here, we assume the reference frame where the $z$-axis is parallel to the 
momentum transfer $\q = \k_i - \k$ and the $y$-axis is parallel to 
$\k_i \times \k$. 

The hadron tensor for the semi-inclusive scattering is given in its general 
form by bilinear combinations of the 
transition matrix elements of the nuclear weak neutral-current operator 
$J^{\mu}$ between
the initial state $|\Psi_0\rangle$ of the nucleus, of energy $E_0$, and the 
final states of energy $E_{\textrm {f}}$, 
given by the product of a discrete (or continuum) state $|n\rangle$ of the residual
nucleus and a scattering state $\chi^{(-)}_{\p_{\mathrm N}}$ of the emitted 
nucleon, with momentum $\p_{\mathrm N}$ and energy $E_{\mathrm N}$. It can be
written as
\begin{eqnarray}
W^{\mu\nu}(\omega,q)  =  
 \sum_{n}  \langle n;\chi^{(-)}_{\p_{\mathrm N}} 
\mid J^{\mu}(\q) \mid \Psi_0\rangle 
\langle 
\Psi_0\mid J^{\nu\dagger}(\q) \mid n;\chi^{(-)}_{\p_{\mathrm N}}\rangle \ 
 \delta (E_0 +\omega - E_{\textrm {f}}) \ ,
\label{eq.ha1}
\end{eqnarray}
where the sum runs over all the states of the residual nucleus.  

Here, the transition matrix elements are calculated in the first
order perturbation theory and in the impulse approximation, i.e., the
incident neutrino interacts with only one nucleon, while the other ones behave 
as spectators. Thus, the transition amplitude is assumed to be adequately 
described as the sum of terms similar to those appearing in the electron
scattering case \cite{book,meucci1}
\begin{eqnarray}
\langle 
n;\chi^{(-)}_{\p_{\mathrm N}}\mid J^{\mu}(\q) \mid \Psi_0\rangle 
= 
\langle\chi^{(-)}_{\p_{\mathrm N}}\mid   j^{\mu}
(\q)\mid \varphi_n \rangle  \ , \label{eq.amp}
\end{eqnarray} 
where 
$\varphi_n = \langle n | \Psi_0\rangle$
describes the overlap between the initial nuclear state and
the final state of the residual nucleus, corresponding to one hole in the 
ground state of the target. 
The single-particle current operator related to the weak neutral current is  
\begin{eqnarray}
  j^{\mu} =  F_1^{\textrm V}(Q^2) \gamma ^{\mu} + 
             i\frac {\kappa}{2M} F_2^{\textrm V}(Q^2)\sigma^{\mu\nu}q_{\nu}	  
	   - G_{\textrm A}(Q^2)\gamma ^{\mu}\gamma ^{5}   +
	     F_{\textrm P}(Q^2)q^{\mu}\gamma ^{5} ,
	     \label{eq.nc}
\end{eqnarray}
where $\kappa$ is the anomalous part of 
the magnetic moment, $q^{\mu} = (\omega , \q)$, $Q^2 = |\q|^2 - \omega^2$, 
is the four-momentum transfer, and
$\sigma^{\mu\nu}=\left(i/2\right)\left[\gamma^{\mu},\gamma^{\nu}\right]$.
$F_1^{\textrm V}$ and $F_2^{\textrm V}$ are the isovector Dirac and Pauli 
nucleon form factors, taken from Ref. \cite{mmd}. 
$G_{\textrm A}$ is the axial form factor and $F_{\textrm P}$ is the  
induced pseudoscalar form factor, which does not contribute to massless 
neutrino scattering.
The vector form factors $F_i^{\mathrm V}$ can be expressed in terms of the 
corresponding electromagnetic form factors for protons $(F_i^{\mathrm p})$ and 
neutrons $(F_i^{\mathrm n})$, plus a possible isoscalar strange-quark 
contribution $(F_i^{\mathrm s})$, i.e.,
\begin{eqnarray}
F_i^{\mathrm V} = \left(\frac{1}{2} - 2\sin^2{\theta_{\mathrm W}}\right)
 F_i^{\mathrm p} -\frac{1}{2} F_i^{\mathrm n} 
 - \frac{1}{2} F_i^{\mathrm s} \ 
 \ \ \ \ ({\mathrm {proton\ knockout}}) \ , \nonumber \\
 F_i^{\mathrm V} = \left(\frac{1}{2} - 2\sin^2{\theta_{\mathrm W}}\right)
 F_i^{\mathrm n} -\frac{1}{2} F_i^{\mathrm p} 
 - \frac{1}{2} F_i^{\mathrm s} \ 
 \ \ \ \ ({\mathrm {neutron\ knockout}}) \ , 
\end{eqnarray}
where $\theta_{\mathrm W}$ is the Weinberg angle $(\sin^2{\theta_{\mathrm W}} 
\simeq 0.2313)$. Standard dipole parametrization is adopted for the 
vector form factors \cite{mmd,hor}. 
The axial form factor is expressed as \cite{mmd}
\begin{eqnarray}
G_{\mathrm A} &=& \frac{1}{2} \left( g_{\mathrm A} - g^{\mathrm s}_{\mathrm
A}\right) G\ \ \ \ ({\mathrm {proton\ knockout}}) \ , \nonumber \\
G_{\mathrm A} &=& -\frac{1}{2} \left( g_{\mathrm A} + g^{\mathrm s}_{\mathrm
A}\right) G\ \ \ \ ({\mathrm {neutron\ knockout}}) \ ,
\end{eqnarray}
where $g_{\mathrm A} \simeq 1.26$, 
$g^{\mathrm s}_{\mathrm A}$ describes possible strange-quark contributions, and 
$G = (1+Q^2/M_{\mathrm A}^2)^{-2}$ with $M_{\mathrm A}$ = 1.026 GeV.
The strange vector form factors are taken as \cite{gar93}
\begin{eqnarray}
F_2^{\mathrm s}(Q^2) & = & \frac {F_2^{\mathrm s}(0)} {(1+\tau) 
(1+Q^2/M_{\mathrm V}^2)^2} , \nonumber \\
F_1^{\mathrm s}(Q^2) & = & \frac {F_1^{\mathrm s} Q^2}{(1+\tau) 
(1+Q^2/M_{\mathrm V}^2)^2} ,
\label{eq.sform}
\end{eqnarray}
where $\tau = Q^2/4M_p^2$, $F_2^{\mathrm s}(0) = \mu_{\mathrm s}$, 
$F_1^{\mathrm s} = -\langle r^2_{\mathrm s}\rangle /6$, and $M_{\mathrm V}$ =
0.843 GeV. The quantity $\mu_{\mathrm s}$ is the strange magnetic moment and 
$\langle r^2_{\mathrm s}\rangle$ the squared {\lq\lq strange radius\rq\rq} 
of the nucleus. 

The differential cross section for the quasielastic neutral-current 
$\nu$($\bar\nu$)-nucleus scattering is obtained from the contraction between 
the lepton and hadron tensors, as in Ref. \cite{Walecka}. 
After performing an integration over the solid angle of the 
final nucleon, we have
\begin{eqnarray}
 \frac{\diff \sigma} {\diff \varepsilon \diff \Omega \diff {\mathrm {T_N}}} = 
 \frac{G^2} {2 \pi^2} \, \varepsilon^2\cos^2 \frac {\vartheta}{2} 
  \Big [ v_0 R_{00} + v_{zz} R_{zz}  - v_{0z} R_{0z} + v_T R_T 
  \pm v_{xy} R_{xy} \Big] \frac {|\p_{\mathrm N}| E _{\mathrm N}} 
 {(2 \pi)^3}\ ,
\label{eq.cs}
\end{eqnarray} 
where $\vartheta$ is the lepton scattering angle and $E _{\mathrm N}$ the
relativistic energy of the outgoing nucleon.
The coefficients $v$ are given as
\begin{eqnarray}
v_0 &=& 1 \quad\quad , \quad\quad
v_{zz} = \frac {\omega^2} {|\q|^2}\ , \nonumber \\
v_{0z}&=& \frac{\omega} {|\q|}  \quad , \quad\quad
v_T=\tan^2\frac {\vartheta}{2} + \frac{Q^2} {2|\q|^2} \ , \nonumber \\
v_{xy}&=& \tan \frac {\vartheta}{2} \left[ \tan^2\frac {\vartheta}{2} +
 \frac{Q^2} {|\q|^2} \right]^{\frac {1} {2}}  \ , 
\label{eq.lepton}
\end{eqnarray}
where the neutrino mass has been neglected. 

The response functions $R$ are given in terms of the components of the 
hadron tensor as
\begin{eqnarray}
R_{00} &=& \int \diff \Omega_{\mathrm N}  \ W^{00}\ , \nonumber \\
R_{zz} &=& \int \diff \Omega_{\mathrm N}  \ W^{zz}\ , \nonumber \\
R_{0z} &=& \int \diff \Omega_{\mathrm N}  \ 2 \ \mathrm{Re} 
(W^{0z})\ , \nonumber \\
R_T  &=& \int \diff \Omega_{\mathrm N}  \ (W^{xx} + W^{yy})\ , 
\nonumber \\
R_{xy} &=&\int \diff \Omega_{\mathrm N}  \ 2\ 
\mathrm{Im}(W^{xy})\ .
\label{eq.rf}
\end{eqnarray}

Here, we are interested
in the single differential cross section with respect to the outgoing nucleon
kinetic energy $\mathrm {T_N}$, i.e.
\begin{eqnarray}
\frac {\diff \sigma} {\diff {\mathrm {T_N}}}  &=& \int 
\left( \frac{\diff \sigma} {\diff \varepsilon \diff \Omega \diff 
{\mathrm {T_N}}}
\right) \diff \varepsilon \diff \Omega  \ , \label{eq.cst} 
\end{eqnarray}
and in the {\lq\lq total\rq\rq} cross section
\begin{eqnarray}
\sigma &=& \int \left(\frac {\diff \sigma} {\diff {\mathrm {T_N}}} \right) 
\diff {\mathrm {T_N}} \ . 
\end{eqnarray}

In the calculation of the transition amplitudes of Eq. \ref{eq.amp} the 
relativistic single-particle scattering wave function 
is written as in Refs. \cite{ee,meucci1} in terms of its upper
component, following the direct Pauli reduction scheme, i.e.,
\inieq
 \chi_{\p_\mathrm{N}}^{(-)} =  \left(\begin{array}{c} 
{\displaystyle \Psi_{\textrm {f}+}} \\ 
\frac{\displaystyle 1} {\displaystyle 
M+ E +S^{\dagger}(E)-V^{\dagger}(E)}
{\displaystyle \mbox{\boldmath $\sigma$}\cdot\p
        \Psi_{\textrm {f}+}} \end{array}\right) \ ,
\fineq
where $S(E)$ and $V(E)$ are the scalar and vector 
energy-dependent
components of the relativistic optical potential for a nucleon
with energy $E$ \cite{chc}. 
The upper component, $\Psi_{\textrm {f}+}$, is related to a two-component
spinor, $\Phi_{\textrm{f}}$, which solves a
Schr\"odinger-like equation containing equivalent central and 
spin-orbit potentials and which is obtained from the relativistic scalar and 
vector potentials \cite{clark,HPa}, i.e.,
\inieq
\Psi_{\textrm {f}+} &=& \sqrt{D^{\dagger}(E)}\ \Phi_{\textrm{f}} \ , 
\nonumber \\ D(E) &=& 1 + \frac{S(E)-V(E)}{M + E} \ , 
\label{eq.darw}
\fineq
where $D(E)$ is the Darwin factor. 

The single-particle overlap functions $\varphi_n$ in Eq. \ref{eq.amp} are taken as 
the Dirac-Hartree solutions of a relativistic Lagrangian
containing scalar and vector potentials \cite{adfx,lala}.

\section{The results for $^{12}$C}
\label{results}

The calculations have been performed for the $^{12}$C nucleus  with the same
bound state wave functions and optical potentials as in
Refs.~\cite{cc,meucci1,meucci2,meucci3,ee}, where the relativistic approach 
was developed to 
study $(\nu_{\mu},\mu^-)$, $\left(e,e^{\prime}p\right)$, 
$\left(\gamma,p\right)$, and $\left(e,e^{\prime}\right)$ reactions.

The relativistic bound state wave functions have been obtained from 
Ref. \cite{adfx}, where relativistic Hartree-Bogoliubov equations are solved in
the context of a relativistic mean-field theory and reproduce
single-particle properties of several spherical and deformed nuclei \cite{lala}. 
The scattering state is computed by means of
the energy-dependent and A-dependent EDAD1 complex phenomenological optical 
potential of Ref. \cite{chc}, which is fitted to proton
elastic scattering data on several nuclei in an energy range up to 1040 MeV.
The initial states $\varphi_n$ are taken as
single-particle one-hole states in the target, with a unitary spectral strength. 
The sum runs over all the occupied states.
In this way we include the contributions of all the nucleons in the nucleus, 
but disregard effects of correlations. These effects, however, are expected to 
be small on the semi-inclusive cross section, and, moreover, should not change
the effects of FSI and of the strange-quark content of the form factors, which
represent the main aim of the present investigation.

\begin{figure}[h]
\begin{center}
\vskip 2.3cm
\includegraphics[height=10cm, width=9cm]{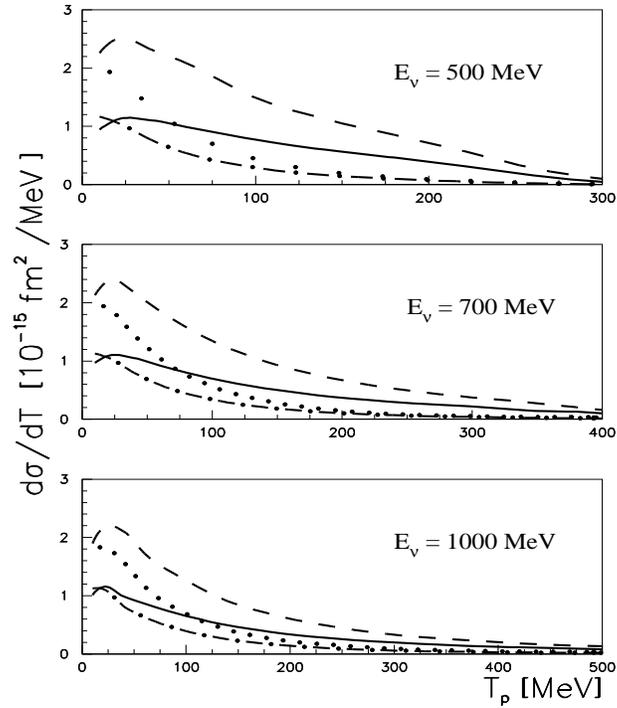} 
\vskip -0.3cm
\caption {Differential cross sections of the $\nu (\bar \nu)$ quasielastic
scattering on $^{12}$C
as a function of the outgoing proton kinetic energy
$T_p$. Solid and dashed lines are the results in RDWIA and RPWIA,
respectively, for an incident neutrino. Dot-dashed and dotted lines are the 
results in RDWIA and RPWIA, respectively, for an incident antineutrino. The 
strangeness contribution is here neglected. }
\label{f.fig1}
\end{center}
\end{figure}

\begin{figure}[h]
\begin{center}
\includegraphics[height=9cm, width=9cm]{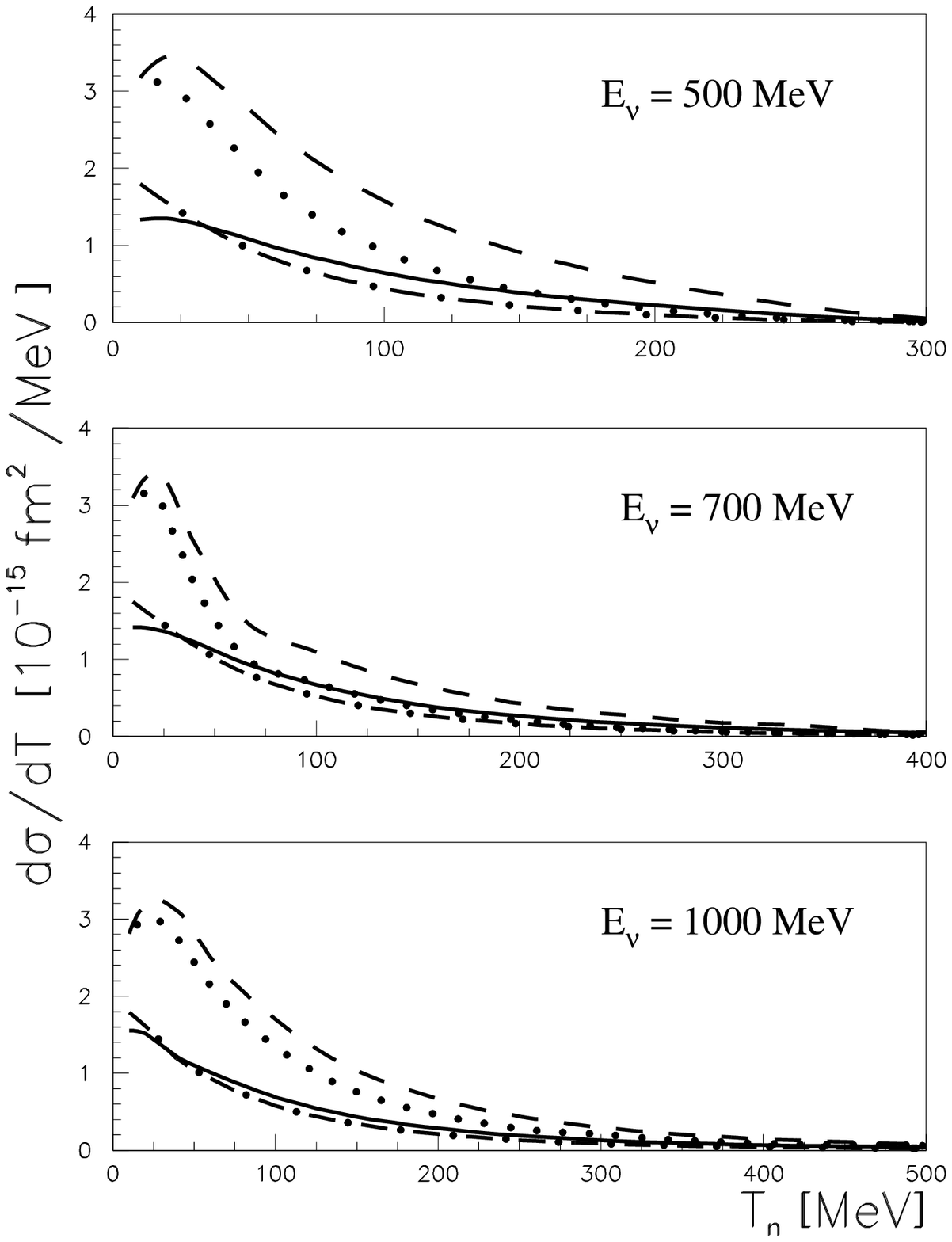} 
\vskip -0.3cm
\caption {The same as in Fig. \ref{f.fig1} but for neutron knockout. }
\label{f.fig2}
\end{center}
\end{figure}

Calculations have been performed for three different values of the incoming 
$\nu$($\bar\nu$) energy, i.e., $E_{\nu (\ove{\nu})} =$ 500, 700, and 1000 MeV. 
Firstly, we investigate the effects of the FSI between the emitted nucleon and 
the residual nucleus, which are described in our approach by means of a  
phenomenological optical-model potential. The cross sections calculated in 
RDWIA and in RPWIA are compared in Fig. \ref{f.fig1} for proton emission and 
in Fig. \ref{f.fig2} for neutron emission.
A reduction of the cross sections of $\, \simeq$ 50\% is found in 
Fig. \ref{f.fig1}, similar to the one obtained in the case of electromagnetic 
proton knockout \cite{book}. However, a larger reduction ($\simeq$ 60\%) 
is obtained in Fig. \ref{f.fig2} when  a neutron 
is emitted. Therefore, the ratio between the cross sections for proton and 
neutron emission is enhanced by distortion.
The comparison between the cross sections for an incident neutrino and
antineutrino gives the
contribution of the axial component of the neutral current. 
The results are in reasonable agreement with those of Ref. \cite{vdv}.

\begin{figure}[h]
\begin{center}
\includegraphics[height=10cm, width=9cm]{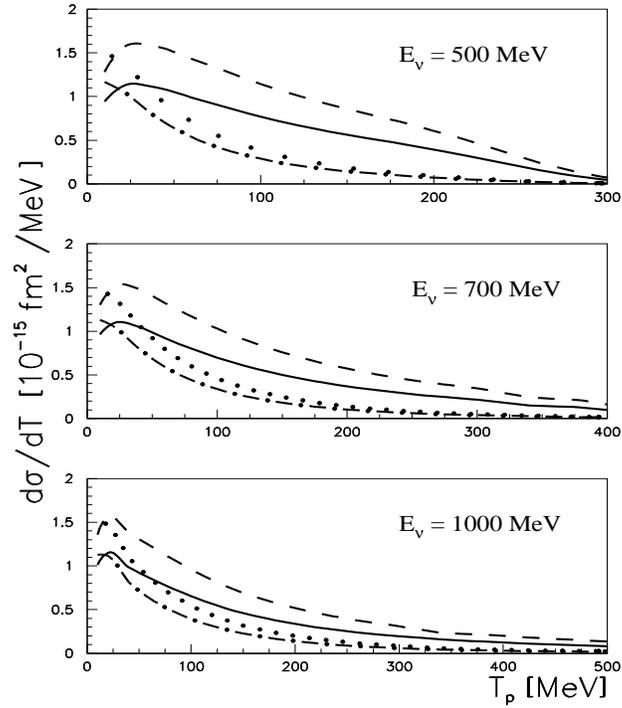} 
\vskip -0.3cm
\caption {Differential cross section in RDWIA of the $\nu (\bar \nu)$ 
quasielastic scattering on $^{12}$C as a function of the outgoing proton 
kinetic energy
$T_p$. Solid and dashed lines are the results with 
$g^{\mathrm s}_{\mathrm A} = 0$ and $g^{\mathrm s}_{\mathrm A} = -0.19$ in 
the case of an incident neutrino. 
Dot-dashed and dotted lines are the corresponding results 
for an incident antineutrino.
}
\label{f.fig3}
\end{center}
\end{figure}

\begin{figure}[h]
\begin{center}
\includegraphics[height=9cm, width=9cm]{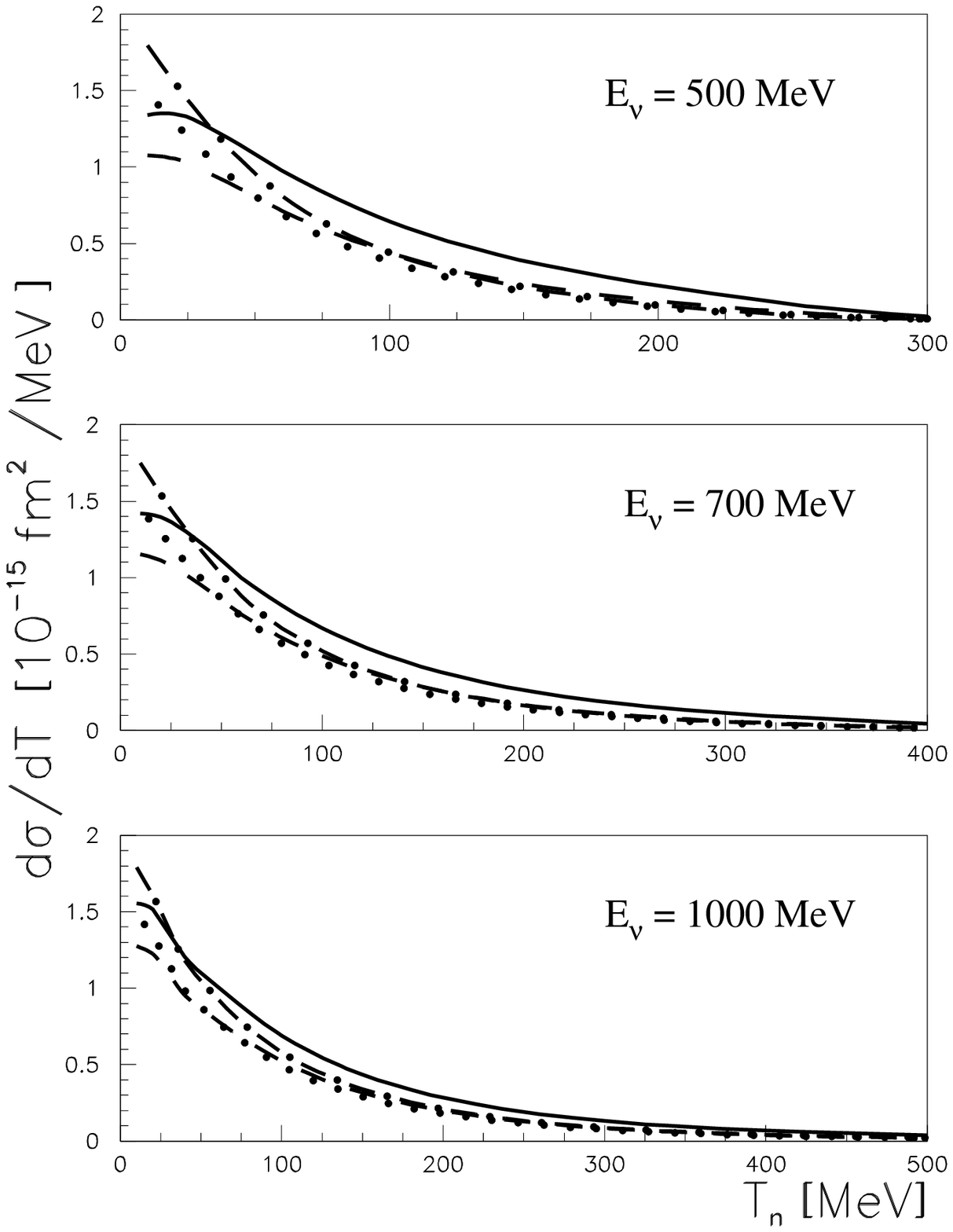} 
\vskip -0.3cm
\caption {The same as in Fig. \ref{f.fig3} but for neutron knockout.
}
\label{f.fig4}
\end{center}
\end{figure}

The effect of a strange-quark contribution to the axial-vector form factor 
is studied in Figs. \ref{f.fig3} and \ref{f.fig4}, where the cross sections 
with $g^{\mathrm s}_{\mathrm A} = -0.19$ are compared with the
corresponding ones 
without any strange-quark content in the weak current.  
The results with  $g^{\mathrm s}_{\mathrm A} = -0.19$ are enhanced by 
$\, \simeq 50$\% with respect to those with 
$g^{\mathrm s}_{\mathrm A} = 0$ in the case of proton knockout and reduced by 
$\, \simeq 30$\% in the case of neutron knockout. 
This different behavior is due to the different sign of the interference term 
$g_{\mathrm A}g^{\mathrm s}_{\mathrm A}$ in the proton and in the neutron 
case. The effect of 
$g^{\mathrm s}_{\mathrm A}$ is less significant in the cross sections with
an incident antineutrino.

\begin{figure}[h]
\begin{center}
\includegraphics[height=9cm, width=8cm]{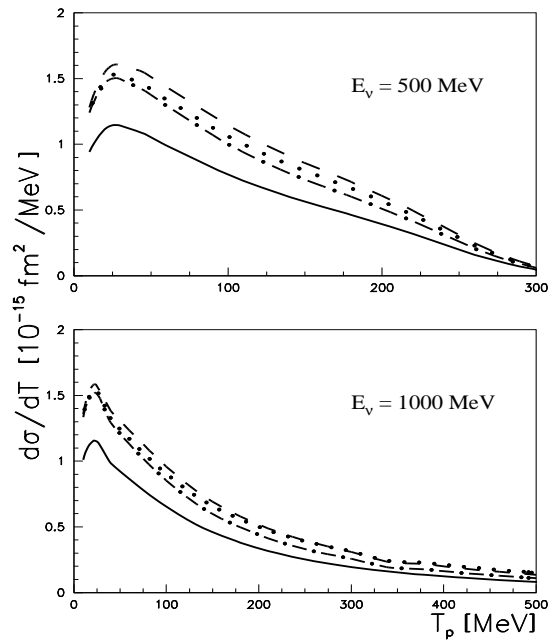} 
\vskip -0.3cm
\caption {Differential cross section of the $\nu$ quasielastic
scattering on $^{12}$C as a function of the outgoing proton 
kinetic energy $T_p$. Solid lines are the results with no strangeness 
contribution, 
dashed lines with $g^{\mathrm s}_{\mathrm A} = -0.19$, dot-dashed lines
with $g^{\mathrm s}_{\mathrm A} = -0.19$ and $ F_2^{\mathrm s}(0) = -0.40$,
dotted lines with $g^{\mathrm s}_{\mathrm A} = -0.19$, 
$ F_2^{\mathrm s}(0) = -0.40$ and
$F_1^{\mathrm s} =- \langle r^2_{\mathrm s}\rangle /6 = 0.53$ GeV$^{-2}$. 
}
\label{f.fig9}
\end{center}
\end{figure}

The effect of the strange vector form factors $F_2^{\mathrm s}(Q^2)$ and
$F_1^{\mathrm s}(Q^2)$ on the 
neutrino quasielastic cross sections is shown in Fig. \ref{f.fig9}. 
We have chosen as in Ref. \cite{gar} $ F_2^{\mathrm s}(0) = -0.40$ and
$F_1^{\mathrm s} =-\langle r^2_{\mathrm s}\rangle /6  = 0.53$ GeV$^{-2}$. 
The $Q^2$ dependence is given in Eq. \ref{eq.sform}.
While $F_2^{\mathrm s}$ decreases the cross sections, $F_1^{\mathrm s}$ gives 
an enhancement that, at $E_\nu = 1$ GeV, almost cancels the effect of 
$F_2^{\mathrm s}$. These effects are smaller than that produced by the strange
axial-vector form factor, $g^{\mathrm s}_{\mathrm A}$, which is shown also in 
Fig. \ref{f.fig9} for a comparison.

\begin{figure}[h]
\begin{center}
\includegraphics[height=10.cm, width=9cm]{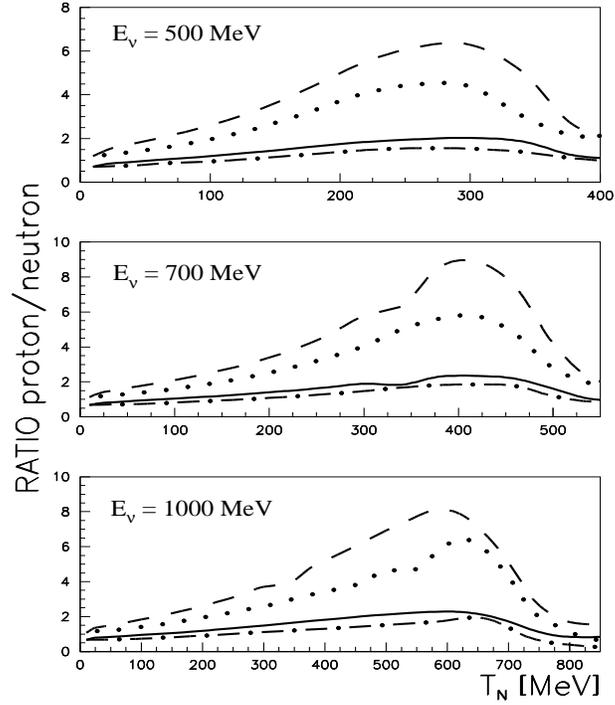} 
\vskip -0.3cm
\caption {Ratio of proton to neutron cross sections of the $\nu$
quasielastic scattering on $^{12}$C as a function of the
outgoing nucleon kinetic energy
$T_N$.
Solid and dashed lines are the results in RDWIA with 
$g^{\mathrm s}_{\mathrm A} = 0$ and $g^{\mathrm s}_{\mathrm A} = -0.19$. 
Dot-dashed and dotted lines are the same results but in PWIA.}
\label{f.fig5}
\end{center}
\end{figure}

In order to separate the effects of the strange-quark contribution and of FSI 
on the cross sections, it was suggested in Refs.\cite{gar92,hor,alberico} to 
measure the ratio of proton to neutron yields, as this ratio is
expected to be less sensitive to distortion effects than the cross sections
themselves. Moreover, from the experimental point of view the ratio is less 
sensitive to the uncertainties in the determination of the incident neutrino 
flux. In Fig. \ref{f.fig5} the ratio for an incident neutrino is
displayed as a function of the outgoing nucleon kinetic energy.
It is very sensitive to $g^{\mathrm s}_{\mathrm A}$ and exhibits, for 
$g^{\mathrm s}_{\mathrm A} = -0.19 $, a pronounced maximum at 
$T_N \simeq 0.6 \ E_\nu$. The RPWIA results are similar to those of 
Ref. \cite{vdv}. A sizable enhancement of the ratio is produced by
FSI. This result is mainly due to the different effects of the optical 
potential in proton and neutron knockout. It means that the argument of looking
for the strange-quark content in this ratio is strengthened by distortion.

\begin{figure}[t]
\begin{center}
\includegraphics[height=9cm, width=9cm]{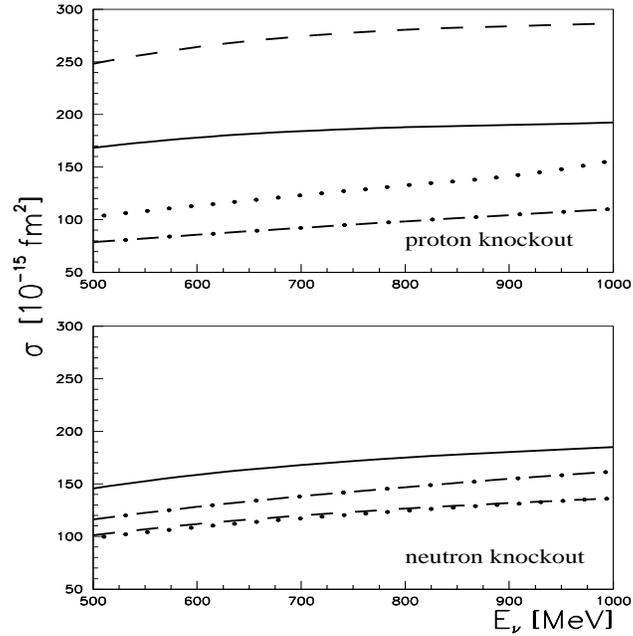} 
\vskip -0.3cm
\caption {The total cross section of the $\nu (\bar \nu)$
quasielastic scattering on $^{12}$C integrated over the outgoing nucleon 
kinetic energy. Upper panel: proton knockout. Lower panel: neutron knockout. 
Line convention as in Fig. \ref{f.fig3}.
}
\label{f.fig6}
\end{center}
\end{figure}

\begin{figure}[h]
\begin{center}
\includegraphics[height=8cm, width=9cm]{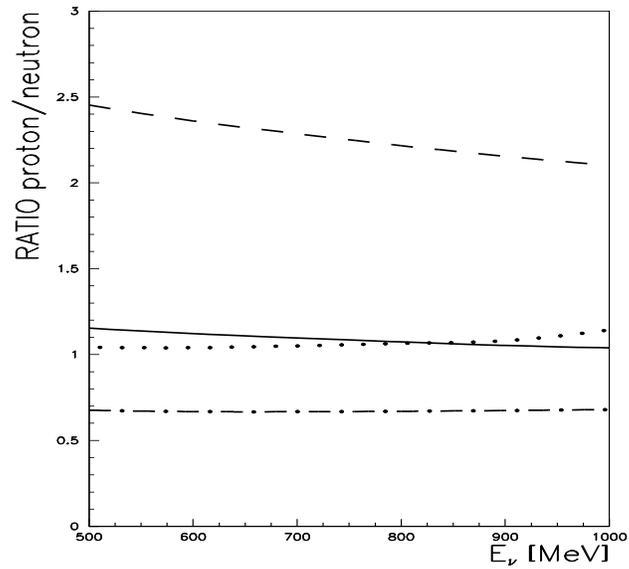} 
\vskip -0.3cm
\caption {Ratio of proton to neutron total cross sections of the $\nu (\bar \nu)$
quasielastic scattering on $^{12}$C as a function of the
incident neutrino (antineutrino) energy. 
Solid and dashed lines are the results in RDWIA with 
$g^{\mathrm s}_{\mathrm A} = 0$ and $g^{\mathrm s}_{\mathrm A} = -0.19$. 
Dot-dashed and dotted lines are the same results but for an incident
antineutrino.}
\label{f.fig7}
\end{center}
\end{figure}

The global effect of the strange-quark contribution is shown in
Fig. \ref{f.fig6}, where the cross sections, integrated over the emitted
nucleon kinetic energy, are displayed. 
The cross section is increased by $\, \simeq 50$\% for proton knockout and 
decreased by $\, \simeq 30$\% for neutron knockout.
The effect is smaller for the cross sections with an incident antineutrino.
The ratio of proton to neutron cross sections is shown in Fig. \ref{f.fig7}.
It is largely enhanced (by a factor $\, \simeq 2$) for neutrino scattering, and 
less enhanced ($\simeq 50$\%) for antineutrino scattering.

\begin{figure}[t]
\begin{center}
\vskip -1.3cm
\includegraphics[height=10cm, width=9cm]{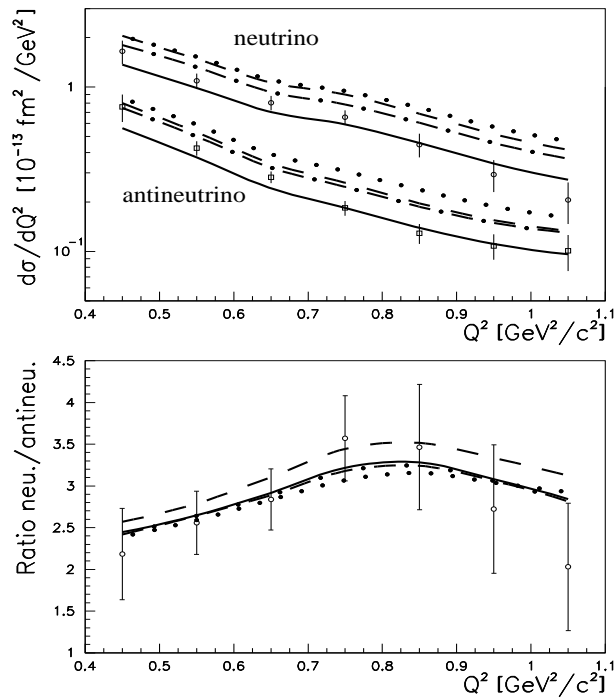} 
\vskip -0.3cm
\caption {Upper panel: differential cross sections of the $\nu (\bar \nu)$
quasielastic scattering, flux-averaged over BNL 
spectrum \cite{bnl}, as a function of the momentum transfer squared. The four 
upper curves
are for incident neutrino and the four lower ones for incident antineutrino.
Lower panel: ratio of neutrino to antineutrino flux-averaged cross sections.
Line convention for both panels as in Fig. \ref{f.fig9}.
Experimental data from Ref. \cite{bnl}.
}
\label{f.fig8}
\end{center}
\end{figure}

Finally, we compare our results with the data of the BNL 734 
experiment \cite{bnl}. Wide band neutrino and antineutrino beams of average
energies of 1.3 and 1.2 GeV, respectively, were used to study 
neutral-current $\nu$($\bar\nu$)-proton scattering. Approximately 80\% of the events
were due to quasielastic scattering from $^{12}$C nuclei and 20\% to elastic
scattering from free protons. Experimental results were presented in the form of
a flux-averaged differential cross section per momentum transfer squared $Q^2$,
which, for a free proton, is related to the recoil energy by the simple relation
$Q^2 = 2 M T_p$. In the case of quasielastic scattering on a complex nucleus, 
this is only approximately true. However, we can still define
\begin{eqnarray}
\langle \frac {\diff \sigma}{\diff Q^2} \rangle = \frac {1}{2M} 
\int \diff T_{\nu (\ove \nu)} \Phi (T_{\nu (\ove \nu)})
\frac {\diff \sigma}{\diff T_p} \ ,
\end{eqnarray}
where the $\nu$($\bar\nu$) spectrum, $\Phi (T_{\nu (\ove \nu)})$, is
normalized to 1. In Fig. \ref{f.fig8} our results for the flux-averaged cross
section are displayed and compared with BNL data. The quasielastic
cross sections 
have been weighed with the free proton scattering ones \cite{gar,beacom}. 
We show results with $g^{\mathrm s}_{\mathrm A} = -0.19$
and without the strange-quark contribution. 
Moreover, following Ref. \cite{gar}, we give the effect of including the
strange vector form factors, $F_1^{\mathrm s}$ and $F_2^{\mathrm s}$, with 
$F_1^{\mathrm s} =-\langle r^2_{\mathrm s}\rangle /6  = 0.53$ GeV$^{-2}$, 
$ F_2^{\mathrm s}(0) = -0.40$, and the $Q^2$ dependence given in 
Eq. \ref{eq.sform}. 
The strange-quark contribution produces an enhancement of the cross sections, 
which makes them slightly higher than the experimental data. The strange weak
magnetic contribution decreases the cross section, while the axial and weak
electric components give an enhancement.
In the lower panel 
of Fig. \ref{f.fig8} the ratio of neutrino to antineutrino 
flux-averaged cross sections is presented and compared with the BNL data. The 
shape of the experimental data is reasonably reproduced.

\section{Summary and conclusions}
\label{conc}

We have presented relativistic calculations for neutral current 
$\nu$($\bar\nu$)-nucleus quasielastic scattering. The reaction mechanism is
assumed to be a direct one: the incident neutrino (antineutrino) interacts with
only one nucleon in the target nucleus and the other nucleons remain as 
spectators. A sum over all single particle
occupied states is performed, using an independent particle model to describe 
the structure of the nucleus. 
The scattering state is an optical-model wave function. 
Results for the $^{12}$C target nucleus have been
presented at neutrino (antineutrino) energies up to 1000 MeV. The optical
potential produces a large reduction of the cross sections. The sensitivity to
the strange-quark content of the axial-vector form factor has been investigated.
A negative value of $g^{\mathrm s}_{\mathrm A} = -0.19$ sensibly increases 
the cross section for proton knockout and decreases it for neutron knockout. 
A large enhancement of the ratio of proton-to-neutron cross sections is 
obtained with respect to the case with $g^{\mathrm s}_{\mathrm A} = 0$. The 
enhancement is increased by FSI. The sensitivity to the strange-quark 
contribution of the vector form factors has also been discussed. 
$F_2^{\mathrm s}$ decreases and $F_1^{\mathrm s}$ enhances the cross 
sections. The two effects tend to cancel each other and are anyhow smaller 
than the effects produced by a strange axial-vector form factor.
Finally, a comparison with the BNL 734 experimental data has been presented. 
The shape of the experimental data is reasonably reproduced.






\end{document}